\documentclass[twocolumn,showpacs,preprintnumbers,amsmath,amssymb, showkeys]{revtex4}

\usepackage{graphicx}
\usepackage{dcolumn}
\usepackage{bm}
\usepackage{epsf}

\begin{document}


\title{Coulomb correlations effects on localized charge relaxation in the coupled quantum dots}

\author{P.\,I.\,Arseyev}
 \altaffiliation{ars@lpi.ru}
\author{N.\,S.\,Maslova}%
 \email{spm@spmlab.phys.msu.ru}
\author{V.\,N.\,Mantsevich}
 \altaffiliation{vmantsev@spmlab.phys.msu.ru}
\affiliation{%
 P.N. Lebedev Physical institute of RAS, 119991, Moscow, Russia\\~\\
 Moscow State University, Department of  Physics,
119991 Moscow, Russia
}%

\date{\today }
11 pages, 10 figures
\begin{abstract}
We analyzed localized charge time evolution in the system of two
interacting quantum dots (QD) (artificial molecule) coupled with the
continuous spectrum states. We demonstrated that Coulomb interaction
modifies relaxation rates and is responsible for non-monotonic time
evolution of the localized charge. We suggested new mechanism of
this non-monotonic charge time evolution connected with charge
redistribution between different relaxation channels in each QD.
\end{abstract}

\pacs{73.63.Kv, 72.15.Lh}
\keywords{D. Electronic transport in quantum dots; D. Relaxation times; D. Non-equilibrium filling numbers}
\maketitle

\section{Introduction}

QDs are unique engineered small conductive regions in the
semiconductor with a variable number of strongly interacting
electrons which occupy well-defined discrete quantum states, for
this reason they are referred to as "artificial" atoms
\cite{Kastner},\cite{Ashoori}. Several coupled QDs form an
"artificial" molecule \cite{Oosterkamp},\cite{Blick_0} and can be
applied for electronic devices creation dealing with quantum
kinetics of individual localized states
\cite{Stafford_0},\cite{Hazelzet},\cite{Cota}. That's why the
behavior of coupled QDs systems in different configurations is under
careful experimental \cite{Waugh},\cite{Blick} and theoretical
investigation \cite{Stafford},\cite{Matveev}. It was demonstrated
experimentally that coupled QDs can vary from the weak tunneling
regime (coupling with the leads is smaller than interaction between
the QDs) to the strong tunneling regime (interaction with the leads
exceeds the QDs coupling) \cite{Oosterkamp},\cite{Livermore}. One of
the most perspective technological goals of QDs integration in a
little quantum circuits deals with careful analysis of
non-equilibrium charge distribution, relaxation processes and
non-stationary effects influence on the electron transport through
the system of QDs
\cite{Angus},\cite{Grove-Rasmussen},\cite{Moriyama},\cite{Landauer},\cite{Loss}.
Electron transport in such systems is governed by Coulomb
interaction between localized electrons and of course by the ratio
between the tunneling transfer amplitudes and the QDs coupling.
Correct interpretation of quantum effects in nanoscale systems gives
an opportunity to create high speed electronic and logic devices
\cite{Tan},\cite{Hollenberg}. So the problem of charge relaxation
due to the tunneling processes between QDs coupled with the
continuous spectrum states in the presence or absence of Coulomb
interaction is really vital. Time evolution of charge states in a
semiconductor double quantum well in the presence of Coulomb
interaction was experimentally investigated in \cite{Hayashi}.
Authors manipulated the localized charge by the initial pulses and
observed pulse-induced tunneling electrons oscillations which were
fitted well by an exponential decay of the cosine function and a
linearly decreasing term. Time dependence of the accumulated charge
and the tunneling current through the single and coupled quantum
wells in the absence of the Coulomb interaction were theoretically
analyzed in \cite{Bar-Joseph}, \cite{Gurvitz_1}, \cite{Gurvitz_2}.
But the authors took into account only two time scales which
determine charge relaxation and neglected the third time scale which
is responsible for charge redistribution between different quantum
wells.

In this paper we consider charge relaxation in a single QD and
double QDs due to the coupling with the continuous spectrum states.
In the case of two coupled QDs tunneling to the continuum is
possible only from one of the QDs. We have found that on-site
 Coulomb repulsion even in one of the dots
results in significant changing of the localized charge relaxation
and leads to formation of several time ranges with strongly
different values of the relaxation rates. We pointed out that the
leading mechanism of non-monotonic charge relaxation is charge
redistribution between the relaxation channels in one of the QDs due
to the Coulomb interaction.

\begin{figure*} [t]
\includegraphics{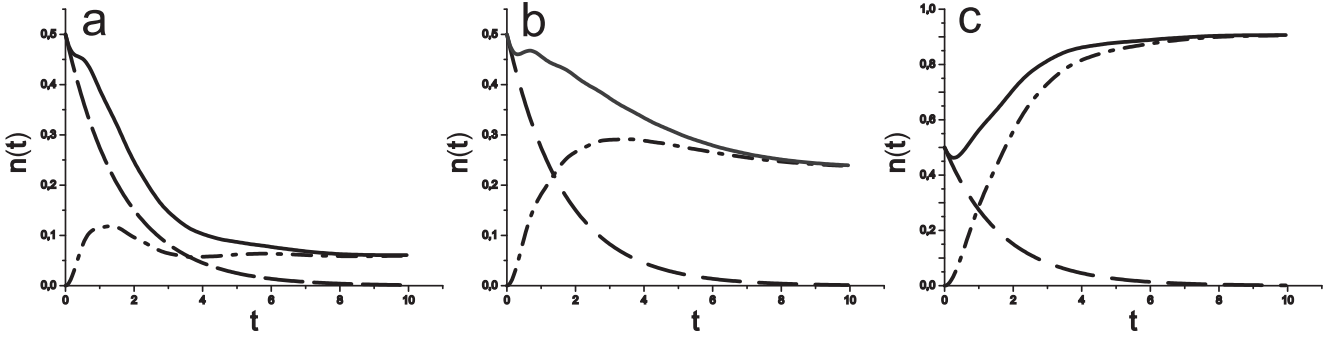}
\caption{Solid line: Localized state filling numbers time evolution
in the QD with energy level $\varepsilon_1$ when distribution
function of the continuous spectrum electrons is taken into account.
Dashed line: Filling numbers time evolution in the QD with energy
level $\varepsilon_1$ without contribution from the conduction
electrons many-particle effects. Dash-dotted line: Filling numbers
evolution only due to the many particle effects caused by the
presence of the distribution function. Tunneling transfer rate
$\gamma_1=0.3$ has the same value for all the figures. a)
$\varepsilon_{1}=1.3$, b) $\varepsilon_{1}=0.3$, c)
$\varepsilon_{1}=-1.3$.}\label{figure1a_1c}
\end{figure*}

\section{Non-stationary tunneling processes in the single QD}

First of all let us consider QD coupled to an electronic reservoir
(conduction electrons states have energies $\varepsilon_k$). We
assume that the single particle level spacing in the dot is larger
than all other energy scales, so that only one, spin-degenerate
level of the QD spectrum is accessible $\varepsilon_1$. Such a
system can be described by the Hamiltonian:

\begin{eqnarray}
\hat{H}&=&\sum_{\sigma}\varepsilon_{1}c_{1\sigma}^{+}c_{1\sigma}+\sum_{k,\sigma}\varepsilon_{k}c_{k\sigma}^{+}c_{k\sigma}+\nonumber\\
&+&\sum_{k,\sigma}T_{k}(c_{k\sigma}^{+}c_{1\sigma}+c_{1\sigma}^{+}c_{k\sigma})
\end{eqnarray}

where $T_{k}$ is a tunneling amplitude between the QD and the
continuous spectrum states which we assume to be independent of
momentum and spin. $c_{1}^{+}/c_{1}$ and $c_{k}^{+}/c_{k}$- electron
creation/annihilation operators in the QD localized state and in the
continuous spectrum states ($k$) correspondingly.

Let us assume that at the initial moment all charge density in the
system is localized in the QD and has the value $n_{1}(0)=n_{0}$. We
shall use Keldysh diagram technique \cite{Keldysh} to describe
charge density relaxation processes in the considered system. Time
evolution of the electron density in the QD is determined by the
Keldysh Green function $G_{11}^{<}$ which is connected with the
localized state filling numbers in the following way:

\begin{eqnarray}
G_{11}^{<}(t,t)=in_{1}(t)
\end{eqnarray}

System of integro-differential equations for the Green function
$G_{11}^{<}(t,t^{'})$ has the form:

\begin{eqnarray}
G_{11}^{0R-1}G_{11}^{<}=\sum_{k}T_{k}G_{k1}^{<}\nonumber\\
G_{k1}^{<}=G_{kk}^{0<}T_{k}G_{11}^{A}+G_{kk}^{0R}T_{k}G_{11}^{<}\nonumber\\
\label{GF}
\end{eqnarray}

and consequently one can obtain the following equation

\begin{eqnarray}
(G_{11}^{0R-1}-\sum_{k}T_{k}^{2}G_{kk}^{0R})G_{11}^{<}=\sum_{k}T_{k}^{2}G_{kk}^{0<}G_{11}^{A}\nonumber\\
\label{green function}
\end{eqnarray}

where continuous spectrum states Green function
$G_{kk}^{0R}(t,t^{'})$ and inverse localized state Green function
$G_{11}^{0R-1}$ in the absence of tunneling processes have the form:

\begin{eqnarray}
G_{kk}^{0R}(t,t^{'})&=&-i\Theta(t-t^{'})e^{-i\varepsilon_{k}(t-t^{'})}\nonumber\\
G_{11}^{0R-1}&=&i\frac{\partial}{\partial t}-\varepsilon_{1}
\end{eqnarray}

In equations (\ref{GF}) and (\ref{green function}) integration over
intermediate time arguments is performed. Finally the solution of
equation (\ref{green function}) can be written as:

\begin{eqnarray}
G_{11}^{<}(t,t)&=&n_{1}(0)e^{-2\gamma_{1}t}+\sum_{k}\int_{0}^{t}\int_{0}^{t}\Theta(t-t_{1})\Theta(t-t_{2})\cdot\nonumber\\
&\cdot&dt_{1}dt_{2}
f(\varepsilon_{k})e^{-i\varepsilon_{k}(t_{1}-t_{2})}\cdot
e^{-i\widetilde{\varepsilon_1}(t-t_{1})}e^{i\widetilde{\varepsilon_1}^{*}(t-t_{2})}\nonumber\\
\label{solution}
\end{eqnarray}

where we define

\begin{eqnarray}
\widetilde{\varepsilon_1}=\varepsilon_{1}-i\gamma_{1}
\end{eqnarray}

and

\begin{eqnarray}
\sum_{k}T_{k}^{2}G_{kk}^{0R}=-i\gamma_1=-i\pi T_{k}^{2}\nu_{k}^{0}
\end{eqnarray}

$\nu_{k}^{0}$-continuous spectrum density of states which is not a
function of energy, $f(\omega)$-Fermi distribution function.

Performing integration in expression (\ref{solution}) and replacing
summation over $k$ by integration over $\omega$ one can get final
expression which describe filling numbers evolution in the quantum
dot due to the interaction with the continuous spectrum states:

\begin{eqnarray}
n_{1}(t)&=&n_{1}(0)\cdot e^{-2\gamma_1t}+\frac{1}{\pi}\int
d\omega\cdot
f(\omega)\frac{\gamma_1}{(\omega-\varepsilon_{1})^{2}+\gamma_{1}^{2}}\cdot\nonumber\\&\cdot&(1+e^{-2\gamma_1t}-2\cos((\omega-\varepsilon_1)\cdot
t)\cdot e^{-\gamma_1t})
\end{eqnarray}

In general, localized charge relaxation law differs from the simple
exponential law even in the absence of Coulomb interaction. Similar
expression was obtained for the initially empty localized states
time evolution by means of Heisenberg equations in
\cite{Bar-Joseph}.

Figure \ref{figure1a_1c} demonstrates the localized state filling
numbers $n_{1}(t)$ time evolution for the different initial
positions of the energy level in the QD. When the continuous
spectrum electrons have Fermi distribution function, charge density
relaxation law strongly differs from the exponential law, especially
when condition $|\varepsilon_1-\varepsilon_F|\leq\gamma_1$ is valid
(solid line in Fig.\ref{figure1a_1c}). This difference can be seen
even when $t\leq\frac{1}{|\varepsilon_1-\varepsilon_F|}$ if
condition $|\varepsilon_1-\varepsilon_F|\gg\gamma_1$ occurs. It is
clearly evident that when contribution from many-particle effects in
the continuous spectrum states is neglected charge relaxation
demonstrates simple exponential law (dashed line in
Fig.\ref{figure1a_1c}). Contribution only from the continuous
spectrum many-particle effects is depicted by the dash-dotted line
in Fig.\ref{figure1a_1c}.

Stationary distribution can be achieved for $t\rightarrow\infty$:

\begin{eqnarray}
n_{1st}=\frac{1}{\pi}\cdot\int d\omega\cdot
f(\omega)\frac{\gamma_1}{(\omega-\varepsilon_1)^{2}+\gamma_{1}^{2}}
\end{eqnarray}

\section{Non-stationary tunneling processes in the system of coupled QDs}

\begin{figure} [t]
\includegraphics{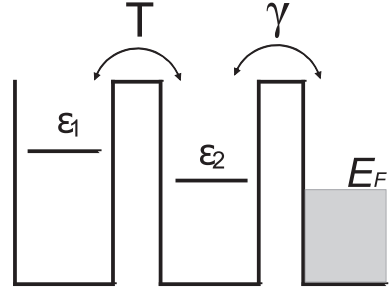}
\caption{Schematic diagram of energy levels in the system of two
coupled QDs. Second QD is also connected with continuous spectrum
states.}\label{figure2}
\end{figure}

Let us now investigate charge relaxation processes in the system of
two coupled QDs with single-electron energy levels $\varepsilon_1$
and $\varepsilon_2$ correspondingly (Fig.\ref{figure2}). QD with
energy level $\varepsilon_2$ is also connected with the continuous
spectrum states. Hamiltonian of the system under investigation has
the form:

\begin{eqnarray}
\hat{H}=\sum_{\sigma}\varepsilon_{1}c_{1\sigma}^{+}c_{1\sigma}+\sum_{\sigma}\varepsilon_{2}c_{2\sigma}^{+}c_{2\sigma}+\sum_{k,\sigma}\varepsilon_{k}c_{k\sigma}^{+}c_{k\sigma}+\nonumber\\
+\sum_{\sigma}T(c_{1\sigma}^{+}c_{2\sigma}+c_{2\sigma}^{+}c_{1\sigma})+\sum_{k,\sigma}T_{k}(c_{k\sigma}^{+}c_{2\sigma}+c_{2\sigma}^{+}c_{k\sigma})
\end{eqnarray}

$T$ and $T_{k}$ are tunneling amplitudes between the QDs and between
the second dot and the continuous spectrum states correspondingly
which we assume to be independent of momentum and spin.
$c_{1}^{+}/c_{1}$($c_{2}^{+}/c_{2}$) and $c_{k}^{+}/c_{k}$-
electrons creation/annihilation operators in the first(second) QD
localized state and in the continuous spectrum states ($k$)
correspondingly.

We assume that at the initial moment all charge density in the
system is localized in the first QD and has the value $n_{1}(0)$.
First of all we have to calculate exact retarded Green functions of
the system. In the absence of tunneling between the QDs Green
functions $G_{11}^{0R}(t-t^{'})$ and $G_{22}^{0R}(t-t^{'})$ are
determined by the expressions:

\begin{eqnarray}
G_{11}^{0R}(t-t^{'})&=&-i\Theta(t-t^{'})e^{-i\varepsilon_1(t-t^{'})}\nonumber\\
G_{22}^{0R}(t-t^{'})&=&-i\Theta(t-t^{'})e^{-i\varepsilon_2(t-t^{'})-\gamma(t-t^{'})}
\end{eqnarray}

where $\gamma=\pi\nu_{k}^{0}T_{k}^{2}$ is the tunneling relaxation
rate from the second QD to the continuous spectrum states.

Retarded electron Green's function $G_{11}^{R}$ yields density of
states in the first QD and can be found exactly from the integral
equation:

\begin{eqnarray}
G_{11}^{R}=G_{11}^{0R}+G_{11}^{0R}T^{2}G_{22}^{0R}G_{11}^{R}
\label{integral_equation}
\end{eqnarray}

The eigenfrequencies $E_{1,2}$ of equation (\ref{integral_equation})
are determined in the following way:

\begin{eqnarray}
(E-\varepsilon_1)(E-\varepsilon_2+i\gamma)-T^{2}=0\nonumber\\
E_{1,2}=\frac{1}{2}(\varepsilon_1+\varepsilon_2-i\gamma)\pm\frac{1}{2}\sqrt{(\varepsilon_1-\varepsilon_2+i\gamma)^{2}+4T^{2}}
\end{eqnarray}

Finally retarded Green's function can be written as:

\begin{eqnarray}
G_{11}^{R}(t,t^{'})=-i\Theta(t-t^{'})(\frac{E_1-\varepsilon_2+i\gamma}{E_1-E_2}e^{-E_{1}(t-t^{'})}-\nonumber\\
-\frac{E_2-\varepsilon_2+i\gamma}{E_1-E_2}e^{-E_{2}(t-t^{'})})
\end{eqnarray}

Let us now analyze time evolution of the electron density in the
considered system. Electron density time evolution is governed by
the Keldysh Green function $G_{11}^{<}(t,t^{'})$ \cite{Keldysh}:

\begin{eqnarray}
G_{11}^{<}(t,t)=in_{1}(t)
\end{eqnarray}

Equation for Green function $G_{11}^{<}$ has the form:

\begin{eqnarray}
G_{11}^{<}(t,t^{'})=G_{11}^{0<}+G_{11}^{0<}T^{2}G_{22}^{0A}G_{11}^{A}+\nonumber\\
+G_{11}^{0R}T^{2}G_{22}^{0R}G_{11}^{<}+G_{11}^{0R}T^{2}G_{22}^{0<}G_{11}^{A}\nonumber\\
\end{eqnarray}

Acting with operator $G_{11}^{0R-1}$ it can be re-written as:

\begin{eqnarray}
G_{11}^{0R-1}G_{11}^{<}(t,t^{'})=(i\frac{\partial}{\partial
t}-\varepsilon_1)G_{11}^{<}(t,t^{'})=\nonumber\\
=T^{2}\int_{0}^{\infty}dt_{1}G_{22}^{0R}(t,t_{1})G_{11}^{<}(t_{1},t^{'})+\nonumber\\
+T^{2}\int_{0}^{\infty}dt_{1}G_{22}^{0<}(t,t_{1})G_{11}^{A}(t_{1},t^{'})
\end{eqnarray}

or in a compact form:

\begin{eqnarray}
(G_{11}^{0R-1}-T^{2}G_{22}^{0R})G_{11}^{<}=T^{2}G_{22}^{0<}G_{11}^{A}
\label{equation}
\end{eqnarray}

Green function $G_{11}^{<}(t,t)$ is determined by the sum of
homogeneous and inhomogeneous solutions. Inhomogeneous solution of
the equation can be written in the following way:

\begin{eqnarray}
G_{11}^{<}(t,t^{'})=T^{2}\int_{0}^{t}dt_{1}\int_{0}^{t^{'}}dt_{2} G_{11}^{R}(t-t_{1})\cdot\nonumber\\
\cdot G_{22}^{0<}(t_1-t_2)G_{11}^{A}(t_2-t^{'})\nonumber\\
\end{eqnarray}

Green function $G_{22}^{0<}$ can be found from equations (\ref{GF})
and (\ref{green function}) substituting index $2$ instead of $1$.

\begin{eqnarray}
G_{22}^{0<}(t,t^{'})=in_{2}(0)\cdot e^{-\gamma
(t+t^{'})}\cdot e^{-i\varepsilon_2(t-t^{'})}+\nonumber\\+\frac{i}{\pi}\int d\omega\cdot f(\omega)\cdot\nonumber\\
\cdot\frac{\gamma}{(\omega-\varepsilon_{2})^{2}+\gamma^{2}}\cdot[e^{-i\omega(t-t^{'})}+e^{-i\varepsilon_2(t-t^{'})-\gamma
(t+t^{'})}\nonumber\\-e^{-i\varepsilon_2t-\gamma t+i\omega
t^{'}}-e^{i\varepsilon_2t^{'}-\gamma t^{'}-i\omega t}] \label{GF_1}
\end{eqnarray}

Green function $G_{22}^{0<}$ contains part with exponential decay,
oscillating term and part which determine stationary solution. In
what follows we'll consider the situation with
$\frac{\varepsilon_i-\varepsilon_F}{\gamma}>>1$ for simplicity. It
means that stationary occupation number in the second QD in the
absence of coupling between QDs is of the order of
$\frac{\gamma}{\varepsilon_2-\varepsilon_F}<<1$. So we can omit
corresponding terms in expression (\ref{GF_1}) for function
$G_{22}^{0<}$.

As we consider initial charge to be localized in the first QD, Green
function $G_{22}^{0<}(0,0)\simeq 0$ and Green function
$G_{11}^{<}(t,t^{'})$ can be determined by the solution of
homogeneous equation. Homogeneous solution of the differential
equation has the form:

\begin{eqnarray}
G_{11}^{<}(t,t^{'})=f_{1}(t^{'})e^{-iE_{1}t}+f_{2}(t^{'})e^{-iE_{2}t}
\label{G_11}
\end{eqnarray}

Since $G^{<}(t,t^{'})$ satisfies the symmetry relation:

\begin{eqnarray}
(G_{11}^{<}(t,t^{'}))^{*}=-G_{11}^{<}(t^{'},t),
\end{eqnarray}

it has the following form:

\begin{eqnarray}
G_{11}^{<}(t^{'},t)=iAe^{-iE_{1}t+iE_{1}^{*}t^{'}}+iBe^{-iE_{1}t+iE_{2}^{*}t^{'}}+\nonumber\\
+iB^{*}e^{-iE_{2}t+iE_{1}^{*}t^{'}}+iCe^{-iE_{2}t+iE_{2}^{*}t^{'}}
\end{eqnarray}

As far as solution has to satisfy homogeneous integro-differential
equation (\ref{equation})(without right hand part), after
substituting expression (\ref{G_11}) to equation (\ref{equation})
one can find the following relation:

\begin{eqnarray}
\frac{f_{1}(t^{'})}{f_{2}(t^{'})}=-\frac{\varepsilon_2-E_{1}-i\gamma}{\varepsilon_2-E_{2}-i\gamma}
\end{eqnarray}

Using the initial condition:

\begin{eqnarray}
G_{11}^{<}(0,0)=in_{1}^{0}
\end{eqnarray}

Time dependence of the filling number $n_{1}(t)$ in the first QD can
be obtained:

\begin{eqnarray}
n_{1}(t)&=&n_{1}^{0}\cdot(A^{'}e^{-i(E_{1}-E_{1}^{*})t}+2Re(B^{'}e^{-i(E_{1}-E_{2}^{*})t})+\nonumber\\
&+&C^{'}e^{-i(E_{2}-E_{2}^{*})t}) \label{filling_numbers_1}
\end{eqnarray}

where coefficients $A'$, $B'$ and $C'$ are equal to:

\begin{eqnarray}
A^{'}&=&\frac{|E_{2}-\varepsilon_1|^{2}}{|E_{2}-E_{1}|^{2}};
C^{'}=\frac{|E_{1}-\varepsilon_1|^{2}}{|E_{2}-E_{1}|^{2}}\nonumber\\
B^{'}&=&-\frac{(E_{2}-\varepsilon_1)(E_{1}^{*}-\varepsilon_1)}{|E_{2}-E_{1}|^{2}}
\label{p1}
\end{eqnarray}

Electron density time evolution in the second QD is determined by
the Green function $G_{22}^{<}(t,t^{'})$ with initial condition
$G_{22}^{<}(0,0)=0$. Green function $G_{22}^{<}(t,t^{'})$ can be
found from equation similar to equation (\ref{equation}) with the
following indexes changing ($1\leftrightarrow2$). Due to the initial
conditions $n_{2}(0)=0$, $n_{1}(0)=n_{0}$, filling numbers in the
second QD $n_{2}(t)$ are determined by the inhomogeneous part of the
solution. Electron filling numbers time dependence in the second QD
$n_{2}(t)$ can be written as:

\begin{eqnarray}
n_{2}(t)&=&(De^{-i(E_{1}-E_{1}^{*})t}+2Re(Ee^{-i(E_{1}-E_{2}^{*})t})+\nonumber\\
&+&Fe^{-i(E_{2}-E_{2}^{*})t}) \label{filling_numbers_2}
\end{eqnarray}

where coefficients $D$, $E$ and $F$ are :

\begin{eqnarray}
D=F=\frac{T^{2}}{|E_{2}-E_{1}|^{2}};\quad
E=-\frac{T^{2}}{|E_{2}-E_{1}|^{2}} \label{p2}
\end{eqnarray}

Expressions (\ref{filling_numbers_1}) and (\ref{filling_numbers_2})
looks like there are three relaxation channels with different time
scales. The first and the second relaxation channels are connected
with relaxation rates $(|E_{1}-E_{1}^{*}|)$ and
$(|E_{2}-E_{2}^{*}|)$ . One more time scale is connected with the
expression $(|E_{1}-E_{2}^{*}|)$. This time scale is responsible for
charge density oscillations in the both QDs, when the following
ratio between $T$ and $\gamma$ is valid: $T/\gamma>1/2$.

In the resonance $\varepsilon_1\simeq\varepsilon_2$ one can find
four different regimes of the system behavior:

1) Realization of the condition {$2T<\gamma$} leads to the absence
of oscillations in the QDs charge density time evolution. In this
case the following expressions are valid:

\begin{eqnarray}
 E_1-E_1^*&=& -i\gamma(1-\sqrt{1-(4T^2)/\gamma^2}) \nonumber \\
 E_2-E_2^*&=& -i\gamma(1+\sqrt{1-(4T^2)/\gamma^2}) \nonumber \\
 E_1-E_2^*&=& -i\gamma \nonumber \\
  && \nonumber
\end{eqnarray}

2) When condition $2T\ll\gamma$ is fulfilled time evolution of the
electron density in the first QD can be described by the expression:

\begin{eqnarray}
 n_1(t)=n_1^0\left[\left(1+\frac{2T^2}{\gamma^2}\right)e^{
 -{{\bf\frac{2T^2}{\gamma}}}t} -
 \frac{2T^2}{\gamma^2}e^{\displaystyle-{{\bf\gamma}}t}
 \right]\nonumber\\
\end{eqnarray}

In this case the main part of the charge decreases with the
relaxation rate

\begin{eqnarray}
\gamma_{res}=2T^2/\gamma
\end{eqnarray}

3) A special regime exists in the system when condition $2T=\gamma$
is valid. Relaxation of the charge in the QDs is non exponential:

\begin{eqnarray}
 n_1(t)=n_1^0(1+\gamma t)e^{\displaystyle-\gamma t}\nonumber\\
n_2(t)=\gamma^{2}t^{2}e^{\displaystyle-\gamma t}
\end{eqnarray}

4) In the case when condition $2T>\gamma$ takes place charge density
oscillations can be seen in the both QDs with the typical frequency
$\Omega=\sqrt{4T^2-\gamma^2}$, for $2T\gg \gamma$:

\begin{eqnarray}
n_1(t)=n_1^0e^{-\gamma t}\,\frac{1}{2}\, \left[1+{\rm
cos}(2Tt)\right] \quad
\end{eqnarray}

Let's now analyze non-resonance case. If we are far from the
resonance, relation $|\varepsilon_1- \varepsilon_2|\gg \gamma,T$
takes place, and the filling numbers relaxation law in the first QD
has the form:

\begin{eqnarray}
n_1(t)=n_1^0\big[\left(1-\frac{2T^2}{(\varepsilon_1-
\varepsilon_2)^2}\right) e^{-{{\bf\frac{2T^2}{(\varepsilon_1-
\varepsilon_2)^2}\gamma}} t} +\nonumber\\+
\frac{2T^2}{(\varepsilon_1- \varepsilon_2)^2}\,{\rm
cos}[(\varepsilon_1- \varepsilon_2)t]\, e^{-{{\bf\gamma}}t} \big]
\end{eqnarray}

Relaxation rates $\gamma_{res}$ and $\gamma_{nonres}$ in the
resonant and non-resonant cases are connected with each other by the
relations:

\begin{eqnarray}
\gamma_{res}=\frac{2T^2}{\gamma} \qquad
\gamma_{nonres}=\gamma_{res}\frac{\gamma^2}{(\varepsilon_1-
\varepsilon_2)^2}
\end{eqnarray}

It is not surprising of course that $\gamma_{res}\gg\gamma_{nonres}$

Let us take into account Coulomb interaction between localized
electrons in the QDs. In this case interaction Hamiltonian can be
written as:

\begin{eqnarray}
H_{int}=U_2n_{2\sigma}n_{2-\sigma}+U_1n_{1\sigma}n_{1-\sigma}
\end{eqnarray}

We shall use self-consistent mean field approximation. It means that
in the derived expressions for the filling numbers time evolution it
is necessary to substitute energy level value $\varepsilon_i$
($i=1,2$) by the expression
$\widetilde{\varepsilon}_i=\varepsilon_i+U\cdot<n_i(t)>$. Then one
should solve self-consistent system of equations for $n_{i}(t)$. We
shall analyze only paramagnetic case:
$n_{i\sigma}=n_{i-\sigma}=n_{i}$.

Such approximation can be applied when the following relations are
fulfilled:

\begin{eqnarray}
|E_{1}-E_{1}^{*}|\ll min(|E_{1}|,|E_{2}|)\nonumber\\
|E_{2}-E_{2}^{*}|\ll min(|E_{1}|,|E_{2}|)\nonumber\\
|E_{1}-E_{2}^{*}|\ll min(|E_{1}|,|E_{2}|) \label{1}
\end{eqnarray}

Inequalities (\ref{1}) mean that one can uncouple rapidly
oscillating Green functions ($G_{11}^{R}$ and $G_{22}^{R}$) and
slowly changing functions $n_{1}(t)$ and $n_{2}(t)$. Suggested
conditions are analogous to the approximations which are used in the
adiabatic approach. In the mean-field approximation the main effect
of Coulomb interaction deals with the detuning changing between
energy levels $\varepsilon_1$ and $\varepsilon_2$. So for simplicity
we shall take into account Coulomb interaction only in the second QD
coupled with the continuous spectrum states. Further we'll
demonstrate that the presence of on-site Coulomb repulsion in both
QDs slightly modifies the obtained results.

Let us discuss the application possibility of the mean-field
approximation in the considered non-stationary case. For the
stationary case when electron filling numbers are changed by
variation of the applied bias or gate voltage in mixed-valence
regime \cite{Anderson} conditions $U/\gamma\geq1$ and
$(\varepsilon_i-\varepsilon_F)/\gamma\geq1$ are important for the
validity of the mean-field approximation. We are interested in the
non-stationary effects, so these conditions are not so crucial. Slow
variations of energy levels due to the localized charge time
evolution allows to obtain reasonable results in the self-consistent
mean-field approximation even in the presence of strong Coulomb
interaction. We also want to point out that initial energy levels
are situated well above the Fermi level
$(\varepsilon_i-\varepsilon_F)/\gamma>>1$, so stationary Kondo
effect can not appear. Energy level in the QD connected with
continuous spectrum is nearly empty. Of course non-zero electron
density appears in the second QD during localized charge relaxation.
So one can try to investigate non-stationary Kondo effect and
estimate the time scale of many-particle correlated state formation.
The simple estimation of this time scale $\tau$ is connected with
the inverse width of Kondo peak.

\begin{eqnarray}
\tau^{-1}\sim\gamma_{Kondo}\sim\sqrt{(\varepsilon_2+U)\cdot\varepsilon_2}\cdot
e^{-\frac{\varepsilon_2\cdot (\varepsilon_2+U)}{U\gamma}}
\end{eqnarray}

Consequently relative values of the system parameters
$U/\varepsilon_2$, $U/\gamma$, $T/\gamma$ and $\varepsilon_2/\gamma$
determine it's behaviour. Our investigations deals with the typical
parameters values demonstrated below on
Fig.\ref{figure4a_4b}-Fig.\ref{figure5a_5b}. For these values of
parameters $\gamma_{Kondo}\sim1\times10^{-2}\gamma$, so
$\tau\sim1\times10^{2}\cdot \tau_{0}$, where $\tau_{0}$ is the
localized charge relaxation time in the second QD due to the
interaction with continuous spectrum. More careful $\tau$ estimation
can be based on the approach suggested in \cite{Glazman}. We can
consider energy level $\varepsilon_2$ position changing due to the
effect of Coulomb interaction $U<n_{2}(t)>$ to be similar to the
influence of time dependent gate voltage on the electron energies in
QDs leading to the spin flip. So following the logic of Glazman et.
al. \cite{Glazman} one can obtain:

\begin{eqnarray}
\tau^{-1}\sim(\gamma\cdot
\frac{T^{2}}{\gamma^{2}}\cdot\frac{U}{\varepsilon_2})^{2}\cdot(\frac{\gamma}{\varepsilon_2})^{2}\cdot(\frac{U}{\varepsilon_2+U})^{2}
\end{eqnarray}

For typical values of system parameters
(Fig.\ref{figure4a_4b}-Fig.\ref{figure5a_5b})
$\tau^{-1}\sim10^{-2}\gamma$.

Thus characteristic time of Kondo peak formation is much larger than
the localized charge relaxation time.

\section{Charge relaxation in the coupled QDs in the presence of on-site Coulomb repulsion}

\begin{figure} [t]
\includegraphics{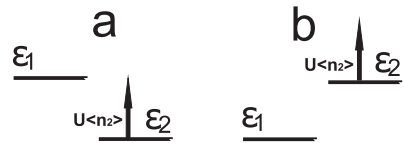}
\caption{Schematic diagram of energy levels position for different
signs of detuning in the presence of Coulomb interaction.}
\label{figure3a_3b}
\end{figure}

\begin{figure} [t]
\includegraphics{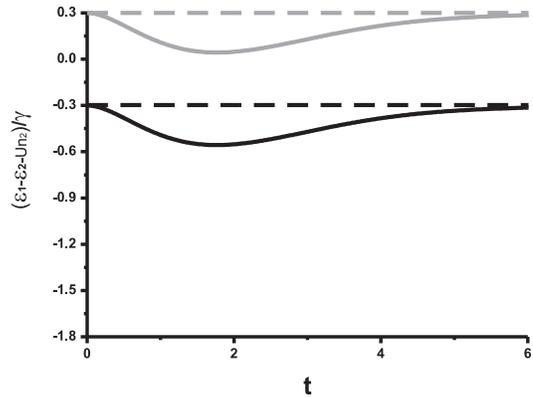}
\caption{Energy levels detuning time evolution. $U/\gamma=0$ and
positive initial detuning
$(\varepsilon_1-\varepsilon_2)/\gamma=(6.3-6.0)/1=0.3$-grey dashed
line; $U/\gamma=0$ and negative initial detuning
$(\varepsilon_1-\varepsilon_2)/\gamma=(6.0-6.3)/1=-0.3$-black dashed
line; $U/\gamma=3$ and positive initial detuning
$(\varepsilon_1-\varepsilon_2)/\gamma=(6.3-6.0)/1=0.3$-grey line;
$U=3$ and negative initial detuning
$(\varepsilon_1-\varepsilon_2)/\gamma=(6.0-6.3)/1=-0.3$-black line.}
\label{figure_3}
\end{figure}

\begin{figure*} [t]
\includegraphics{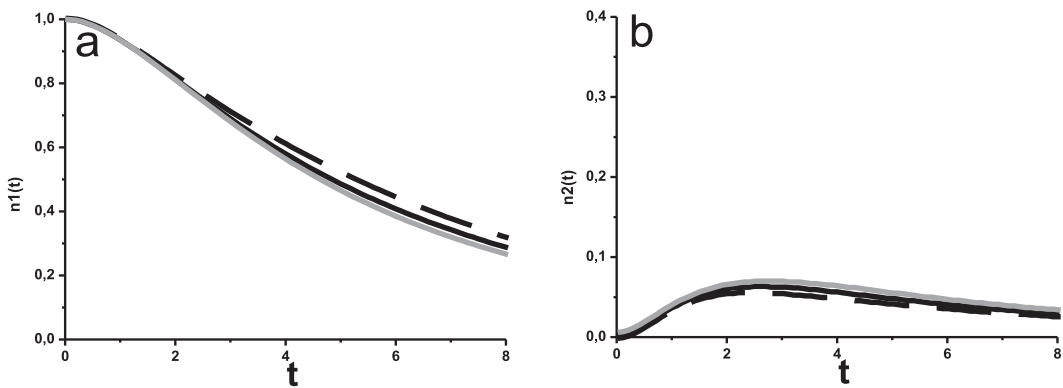}
\caption{Filling numbers evolution in the first a). $n_{1}(t)$ and
second b). $n_{2}(t)$ QDs. $U/\gamma=0$-black line both for positive
$(\varepsilon_1-\varepsilon_2)/\gamma=(6.3-6.0)/1=0.3$ and negative
$(\varepsilon_1-\varepsilon_2)/\gamma=(6.0-6.3)/1=-0.3$ detuning,
$U/\gamma=3$ and
$(\varepsilon_1-\varepsilon_2)/\gamma=(6.3-6.0)/1=0.3$ -grey line,
$U/\gamma=3$ and
$(\varepsilon_1-\varepsilon_2)/\gamma=(6.0-6.3)/1=-0.3$-dashed line.
Parameters $T/\gamma=0.6$ and $\gamma=1.0$ are the same for all the
figures.} \label{figure4a_4b}
\end{figure*}

We start by discussing the case of weak Coulomb interaction when the
ratio $Un_{2}(t)/\gamma\leq\Delta\varepsilon/\gamma$ is fulfilled
(Fig.\ref{figure4a_4b}). If the initial detuning between energy
levels has positive value ($\varepsilon_1>\varepsilon_2$
Fig.\ref{figure3a_3b}a) localized charge relaxation rate in the
first QD and full charge density in the second QD increase
(Fig.\ref{figure4a_4b}a,b grey line) in comparison with the case
when Coulomb interaction is absent (Fig.\ref{figure4a_4b}a,b black
line). Relaxation rate increases due to the decreasing of initial
detuning value $\Delta\varepsilon/\gamma$ caused by Coulomb
interaction (Fig.\ref{figure3a_3b}a, Fig.\ref{figure_3}).
Fig.\ref{figure_3} demonstrates the detuning time evolution and
reveals that in the absence of Coulomb interaction energy levels
detuning has constant value (Fig.\ref{figure_3} grey dashed line).
The presence of on-site Coulomb repulsion results in the smaller
detuning values in the system except the initial time moment and
time period when the system is quite empty and Coulomb interaction
can be neglected (Fig.\ref{figure_3} grey line).

In the opposite case of negative initial energy levels detuning
($\varepsilon_1<\varepsilon_2$ Fig.\ref{figure3a_3b}b) even small
values of Coulomb interaction results in the effective increasing of
energy levels spacing (Fig.\ref{figure_3} black line). Consequently,
relaxation rate in the first QD and full charge density in the
second QD decrease (Fig.\ref{figure4a_4b}a dashed line) in
comparison with the case when Coulomb interaction is absent
(Fig.\ref{figure4a_4b}b dashed line) and detuning takes on constant
value.

\begin{figure*} [t]
\includegraphics {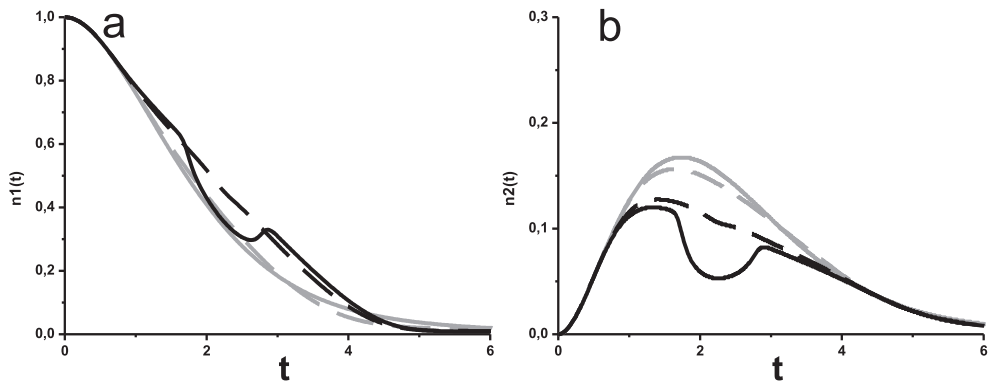}
\caption{Filling numbers evolution in the first a). $n_{1}(t)$ and
second b). $n_{2}(t)$ QDs. $U/\gamma=0$-grey line, $U/\gamma=6$-grey
dashed line, $U/\gamma=12$-black dashed line, $U/\gamma=14$-black
line. Parameters
$(\varepsilon_1-\varepsilon_2)/\gamma=(6.3-6.0)/1=0.3$,
$T/\gamma=0.6$ and $\gamma=1.0$ are the same for all the figures.}
\label{figure5a_5b}
\end{figure*}

\begin{figure} [t]
\includegraphics[width=80mm]{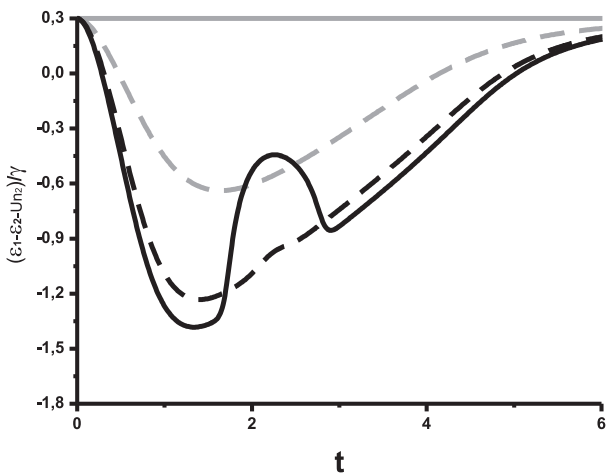}
\caption{Energy levels detuning time evolution in the case of
positive initial detuning
$(\varepsilon_1-\varepsilon_2)/\gamma=(6.3-6.0)/1=0.3$.
$U/\gamma=0$-grey line; $U/\gamma=6$-grey dashed line;
$U/\gamma=12$-black dashed line; $U/\gamma=14$-black.}
\label{figure_5}
\end{figure}

Let us now focus on the charge relaxation processes due to the
presence of strong Coulomb interaction
($Un_{2}(t)/\gamma\geq\Delta\varepsilon/\gamma$) and positive
initial detuning (Fig.\ref{figure5a_5b}). In this case filling
numbers time evolution in the first QD reveals three typical time
intervals with different values of the relaxation rates. The first
one corresponds to the time interval $0<t<t_{02max}$, where
$t_{02max}$- is an instant of time when the filling numbers in the
second QD reach maximum value $n_{2max}$ (Fig.\ref{figure5a_5b}b).
Simultaneously filling numbers in the first QD demonstrate the
relaxation rate changing (bend formation) (Fig.\ref{figure5a_5b}a).
The appropriate detuning behavior is presented on the
Fig.\ref{figure_5}. One can find that the presence of strong on-site
Coulomb repulsion results in fast compensation of initial positive
detuning. Further time evolution leads to the formation of negative
detuning between energy levels and it is clearly evident that the
filling numbers maximum in the second QD corresponds to the maximum
energy levels spacing. This time interval reveals charge relaxation
with the typical rate very close to the $\gamma_{res}$. Coulomb
interaction increasing in this time interval results in the weak
decreasing of relaxation rate in comparison with the case when
Coulomb correlations are neglected.

\begin{figure*} [t]
\includegraphics {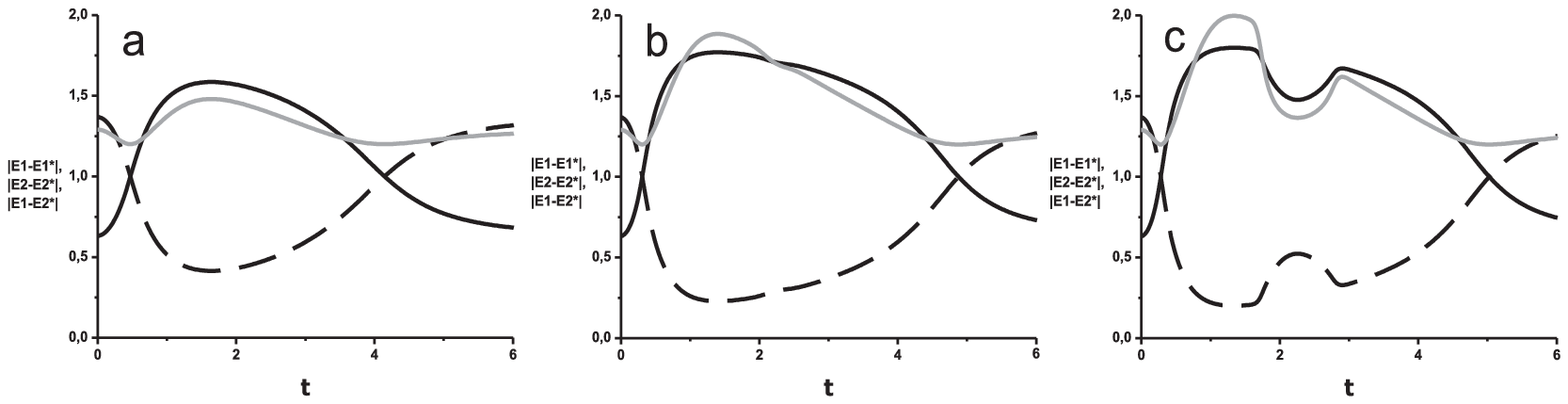}
\caption{Each channel relaxation rates time evolution for the values
of parameters
$(\varepsilon_1-\varepsilon_2)/\gamma=(6.3-6.0)/1=0.3$;
$T/\gamma=0.6$; $\gamma=1.0$. $E_1-E_{1}^{*}$-black line;
$E_2-E_{2}^{*}$-dashed black line; $E_1-E_{2}^{*}$-grey line; a).
$U/\gamma=6$; b). $U/\gamma=12$; c). $U/\gamma=14$.}
\label{figure6a_6c}
\end{figure*}

The next time interval $t_{02max}<t<t_{01min}$ ($t_{01min}$-is an
instant of time when the filling numbers in the first QD $n_{1}(t)$
achieve minimum value $n_{1min}$) reveals localized charge
relaxation with the typical rate very close to the
$\gamma_{nonres}$. Filling numbers time evolution in the first and
second QDs simultaneously demonstrates dip's formation
(Fig.\ref{figure5a_5b}a,b) which corresponds to the local minimum of
the effective detuning (Fig.\ref{figure_5}). This phenomenon can be
explained by the influence of the following physical mechanism:
charge redistribution between the QDs by means of relaxation channel
with the typical relaxation rate $(|E_{1}-E_{2}^{*}|)$
(eq.\ref{filling_numbers_1}) in the presence of strong on-site
Coulomb repulsion. Charge redistribution strongly governs the system
behavior in the second time interval due to the presence of
significant real part of the relaxation rate $(|E_{1}-E_{2}^{*}|)$
which is proportional to the value $Un_{2}(t)$
(Fig.\ref{figure6a_6c} grey line). In the first time interval
relaxation occurs regularly because real part of the relaxation rate
$(|E_{1}-E_{2}^{*}|)$ is negligible or nearly absent due to the
small values of filling numbers in the second QD. Contribution from
the charge redistribution mechanism becomes crucial only when
filling numbers in the second QD reach maximum value. These effect
self-consistently influence the charge dynamics in the proposed
system by means of effective detuning changing.

The third time interval demonstrates charge relaxation with the
typical rate very close to the $\gamma_{res}$. This interval exists
due to the decreasing of the filling numbers amplitude in the second
QD. It leads to the increasing of the relaxation rate value in
comparison with the previous time interval. The presence of on-site
Coulomb repulsion leads to the small detuning values in comparison
with the second time interval due to the fact that the system is
quite empty and Coulomb interaction can be neglected
(Fig.\ref{figure_5}). Consequently charge redistribution also can be
omitted. Therefore with the increasing of time the effective
detuning aspire to the value without any Coulomb correlations in the
system.

Strong Coulomb interaction significantly influence on the relaxation
processes. To analyze the mechanism of relaxation law modification
one have to examine the relaxation exponents evolution, which
determine charge relaxation rates changing in each channel of the
QDs. Moreover we shall analyze time evolution of the preexponential
factors which govern charge re-distribution between the relaxation
channels. This analysis will be carried out for the most interesting
case of positive initial detuning, when Coulomb interaction leads to
the dip's formation.

Let us start from the analysis of the exponents evolution. Their
behavior is just the same for the both QDs (Fig.\ref{figure6a_6c}).
In the absence of Coulomb interaction second channel relaxation rate
$E_2-E_{2}^{*}$ always exceeds the first channel $E_1-E_{1}^{*}$
relaxation rate. This ratio between relaxation rates also takes
place at the initial time moment in the presence of Coulomb
interaction. Coulomb interaction results in the dip and peak
formation in the second and first relaxation channels accordingly
(Fig.\ref{figure6a_6c}a,b). First channel relaxation rate maximum
value corresponds to the second channel relaxation rate minimum
value. For the large time values evolution laws demonstrate constant
values of relaxation rates for both relaxation channels equal to the
values obtained without Coulomb interaction. Splitting of the peak
in the first relaxation channel and dip in the second one can be
seen with further increasing of Coulomb interaction. Moreover peaks
in the first relaxation channel correspond to the dips in the second
relaxation channel and dip in the first relaxation channel
corresponds to the peak in the second relaxation channel
(Fig.\ref{figure6a_6c}c).

\begin{figure*} [t]
\includegraphics{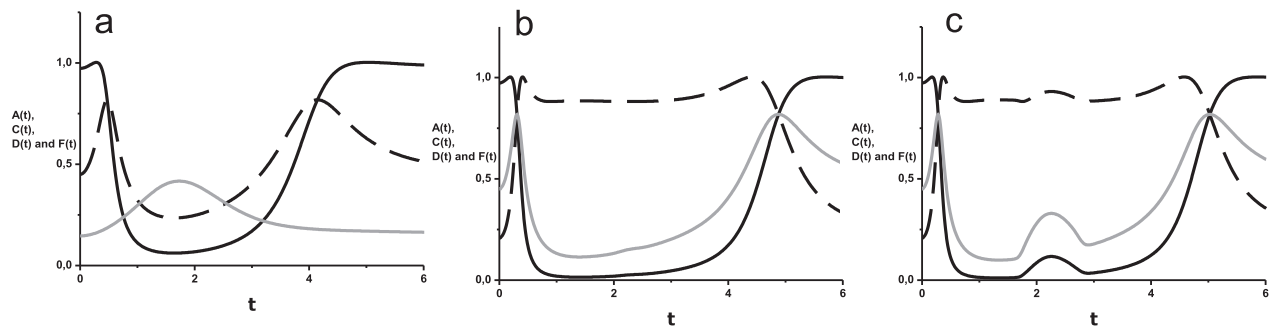}
\caption{Each relaxation channel amplitude time evolution as a
function of time for the values of parameters
$(\varepsilon_1-\varepsilon_2)/\gamma=(6.3-6.0)/1=0.3$;
$T/\gamma=0.6$; $\gamma=1.0$. Preexponential factor for the first
relaxation channel in the first QD $A(t)$-black line; preexponential
factor for the second relaxation channel in the first QD
$C(t)$-dashed black line; preexponential factors for the first and
second relaxation channels in the second QD $D(t)$ and $F(t)$-grey
line. a). $U/\gamma=6$; b). $U/\gamma=12$; c). $U/\gamma=14$.}
\label{figure7a_7c}
\end{figure*}

Let us now focus on the preexponential factors (relaxation channel's
amplitudes) time evolution in the presence of Coulomb interaction.
In the second QD preexponential factors time evolution is determined
by the same law ($D(t)$ and $F(t)$ expression \ref{p2}) (Fig.
\ref{figure7a_7c} grey line). Time evolution of the preexponential
factors in the first QD is quite different ($A(t)$ and $C(t)$
expression \ref{p1}).

First relaxation channel amplitude always exceeds second relaxation
channel amplitude in the first QD and both relaxation channels
amplitudes in the second QD in the absence of Coulomb interaction.
This ratio is also valid at the initial time moment in the case of
Coulomb interaction between localized electrons.

Relaxation channel's amplitudes time evolution in the second QD
($D(t)$ and $F(t)$ in Fig.\ref{figure7a_7c}b) demonstrates maximum
which corresponds to the minimum for the both relaxation channel's
amplitudes in the first QD for small Coulomb interaction values.
With the increasing of Coulomb interaction (Fig.\ref{figure7a_7c}b)
first relaxation channel amplitude in the first QD $A(t)$ and both
relaxation channel's amplitudes in the second QD ($D(t)$ and $F(t)$)
tend to zero, second relaxation channel amplitude in the first QD
$C(t)$ tends to the unity. Further time evolution demonstrates that
all channels amplitudes in both QDs turn to constant values equal to
the values obtained without Coulomb interaction.

Increasing of the Coulomb interaction results in simultaneous peaks
formation for all the relaxation channels (Fig.\ref{figure7a_7c}c).

Comparing the obtained results which describe  exponents evolution
(charge relaxation rates in each channel) and evolution of the
preexponential factors (time evolution of the each relaxation
channel amplitude) one can easily reveal that charge redistribution
between the relaxation channels in the same QD occurs. At the
initial moment most part of the charge is concentrated in the first
relaxation channel of the first QD and later localized charge
redistributes to the second relaxation channel of the first QD. The
following time increasing again demonstrates charge localization in
the first relaxation channel of the first QD. Charge in the second
QD is equally distributed between both relaxation channels.

Charge relaxation in the presence of Coulomb interaction in both
quantum dots is determined by the charge density redistribution
among different channels in the same QD and by the changing of
relaxation rates of each channel. So due to Coulomb interaction the
leading mechanism of non-monotonic charge relaxation in each QD is
charge redistribution between the channels in a separate QD at
particular range of the system parameters.

\begin{figure*} [t]
\includegraphics{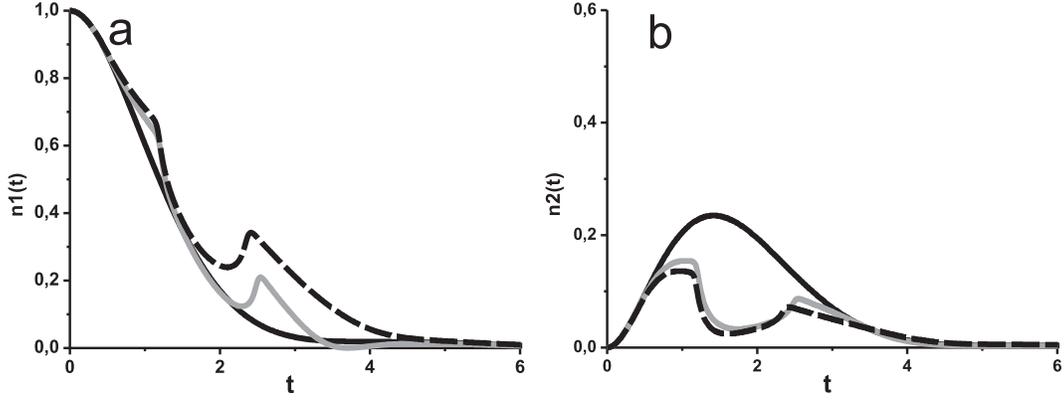}
\caption{Filling numbers evolution in the first a). $n_{1}(t)$ and
second b). $n_{2}(t)$ QDs without Coulomb interaction
($U_{1}/\gamma=0$ and $U_{2}/\gamma=0$-black line), with Coulomb
interaction in the second QD ($U_{1}=0$ and $U_{2}=14$-grey line),
with Coulomb interaction in both QDs ($U_{1}/\gamma=3$ and
$U_{2}/\gamma=14$- black dashed line). Parameters
$(\varepsilon_1-\varepsilon_2)/\gamma=(6.3-6.0)/1=0.3$,
$T/\gamma=0.6$ and $\gamma=1.0$ are the same for all the figures.}
\label{figure8a_8b}
\end{figure*}

The last point of our discussion deals with comparison between
results obtained for suggested model and more natural from
experimental point of view situation when Coulomb interaction is
taken into account in both QDs. Fig.\ref{figure8a_8b} demonstrates
localized charge time evolution for the both relaxation channels.
Grey line corresponds to the case when Coulomb interaction exists
only in the second QD and black-dashed line describes the situation
when Coulomb interaction is taken into account in the both QDs. It
is evident that Coulomb interaction in the both QDs results in more
rapid compensation of energy levels detuning for
$\Delta\varepsilon/\gamma>0$ and doesn't lead to qualitative changes
of the obtained results. So (as it was mentioned above) for
simplicity it is sufficient to consider Coulomb interaction only in
one QD. In any case Coulomb interaction effectively changes level
spacing, which control the charge redistribution between the
relaxation channels in a single QD as well as between two QDs.

\subsection{Conclusion}

We have analyzed time evolution of localized charge in the system of
two interacting QDs both in the absence and in the presence of
Coulomb interaction between localized electrons within a particular
quantum dot. We have found that Coulomb interaction modifies the
relaxation rates and the character of localized charge time
evolution. It was shown that several time ranges with considerably
different relaxation rates arise in the system of two coupled QDs.
We demonstrated that the presence of Coulomb interaction leads to
strong charge redistribution between different relaxation channels
in each QD. So we can conclude that non-monotonic behavior of charge
density is not the result of charge redistribution between the QDs
but is determined by charge redistribution between the relaxation
channels in a single QD.

In any real situation Coulomb interaction is present in both QDs.
But for simplicity it is sufficient to consider Coulomb interaction
only in one QD since the main role of Coulomb interaction is charge
redistribution between the relaxation channels in a single QD as
well as between QDs due to modification of the detuning between
energy levels in QDs.

\section{Acknowledgements}
We acknowledged the financial support from RFBR and Leading
Scientific School grants.


\pagebreak

\end{document}